\documentclass[fleqn,usenatbib]{mnras}

\usepackage[T1]{fontenc}
\usepackage{ae,aecompl}

\usepackage{graphicx}	

\title[Superflare on Proxima Centauri]{Observation of a possible superflare on Proxima Centauri}

\author[J. Kielkopf et al.]{John F. Kielkopf,$^{1,2}$ \thanks{E-mail: kielkopf@louisville.edu}
Rhodes Hart,$^{2}$
Bradley D. Carter,$^{2}$
and Stephen C. Marsden$^{2}$
\\
${1}$University of Louisville, Department of Physics and Astronomy, Louisville, KY 40292, USA \\
${2}$ University of Southern Queensland, Centre for Astrophysics, 
Australia\\
}

\date{Accepted XXX. Received YYY; in original form ZZZ}
\pubyear{2018}

\begin{document}
\label{firstpage}
\pagerange{\pageref{firstpage}--\pageref{lastpage}}
\maketitle


\begin{abstract}
We  report  the  observation on UT 2017.07.01 of  an  unusually  powerful  flare  detected  in 
near-infrared continuum photometry of Proxima Centauri.  During a campaign monitoring the star for possible exoplanet transits, we identified an increase in Sloan i' flux leading to an observed peak at BJD 2457935.996 that was at least 10\% over pre-flare flux in this band. It was followed by a 2-component rapid decline in the first 100 seconds that became a slower exponential decay with time constant of 1350 seconds. 
A smaller flare event 1300 seconds after the first added an incremental peak
flux increase of 1\% of pre-flare flux. Since the onset of the flare was not
fully time-resolved at a cadence of 62 seconds, its actual peak value is unknown but greater than the time-average over a single exposure of 20 seconds. The i' band is representative of broad optical and near-IR continuum flux over which the integrated energy of the flare is  100 times  the  stellar  luminosity. This  meets  the  criteria  that  established  the concept  of  superflares  on  similar  stars.  The resulting implied ultraviolet flux and space weather could have had an extreme effect on the atmospheres of planets within the star's otherwise-habitable zone.

\end{abstract}


\section{Introduction}

In 1859 two independent visual observers of the Sun saw and wrote about a sudden brief white light brightening in a 
sunspot group~\citep{hodgson1859,carrington1859}.  Now known
as the Carrington event, it remains the largest solar flare ever in written history~\citep{schaefer2012}.  Carrington recognized it as ``exceedingly rare''
and noted it was followed after midnight by a magnetic storm on Earth in both hemispheres.  Hodgson said it was of ``dazzling brilliancy'' and
compared it to $\alpha$~Lyrae seen in a large telescope at low magnification.  Modern analysis puts the total energy output of this flare at
$10^{32}$ erg ($10^{25}$ joule).~\citep{schaefer2012,maehara2012,tsurutani2003} The total solar radiant power across the entire spectrum is $10^{26}$ watts, and while the Carrington event lasted for a few
minutes it was still  a significant fraction of the Sun's output especially at its onset.  Since then we have come to recognize and study in great detail the routine
flares on the Sun, and to see them in integrated light from similar solar-type stars, and from cooler dwarfs that show magnetic fields and more frequent low energy flares.

Over a century later in 1989 Schaefer recognized that  there was a class of these exceptional flares, for which two dozen examples of normal stars from types B2~V to M2~I were identified at the time, with even greater brief
flashes of output relative to their star's average luminosity, from  $10^{32}$ erg to some possibly as energetic as $10^{40}$ erg~\citep{schaefer1989}.  Since then more have been found 
with bolometric energies up to $10^{36}$ erg in both ground-based
observations and in data from the Kepler space telescope such as described by \citet{schaefer2000,maehara2012} and \citet{candelaresi2014}. The Kepler data have been analyzed by Maehara et al to identify 365 such
events in a sample of 83,000 stars observed for 120 days, and by Candelaresi et al. to find 1690 superflares in 380 of 117,661 stars. 

Thus, in the context of flares we know to occur more commonly on magnetically active cool stars, including red dwarfs and solar-type stars, the one seen by Carrington and Hodgson was barely exceptional.  However, since occasionally one is detected in
stars very similar to the Sun in which the luminosity of the star increases by such a large amount that life on otherwise-habitable planets would be in jeopardy, it is
reasonable to ask what the probability is of  a solar \textit{super}\/ flare.  Fortunately, at least for the duration of written human history, this has not happened for our star, and we can
set that as low as $10^{-5}$ per year with a large error bar from a sample of one.    We do not yet know with confidence whether such an event is  even possible given the
structure of our star, or if it is inevitable over the life of the Sun. Indeed, it is puzzling that in  the catalog of superflare events and their stars now known,  there are
are so few with extrasolar planets identified by transits~\citep{armstrong2016}.   For that reason it is especially interesting when we find a superflare on a star that is a known host to planets, and on one so close and bright that it can be studied in detail.

Obviously the use of ``super'' to describe these exceptionally rare events implies also exceptional energy output.  The exact threshold for this criterion is a somewhat arbitrary choice that was set when Schaefer made the first identification of a class of unusual rare  events~\citep{schaefer1989,schaefer2000}.  The recent Kepler data studies have revealed a continuum of flare energies and a decreasing probability for the most energetic of them.  There are selection effects in the recent work because flares are identified by threshold crossings in the raw photometric data, and for Kepler at least, the stars are typically faint and distant to begin with.  \citet{candelaresi2014}  use a total energy release of $5\times10^{34}$ erg over a few hours as the boundary above which a flare is super in type G, K and M stars, and note that the largest flares on the Sun are $10^{32}$ erg as in the Carrington and Hodgson event, that is less than 1\% of their threshold energy for superflare status.  It is important to recognize that this in no way says that the Sun cannot have a superflare.  It has been more than a century since the Carrington flare, but there is less than 10,000 years of written history. Even if we include the geological data,  the continuum of flare energies seen in Kepler observations imply that a flare over the ``super'' threshold could occur, while our existence confirms the improbability it being life-threatening. It seems reasonable to use the term superflare for events that are 10 times as energetic as the one Carrington and Hodgson actually saw visually, that is, $10^{33}$ erg is a thoughtful threshold to this unusual regime for solar-type and cooler stars as adopted by \citet{schaefer2000}. 

For the purpose of establishing the significance of unusual flares on exoplanets, a criterion based on peak flux relative to the ambient flux of the star is more revealing. It takes into account the location of the star's habitable zone which identifies planets of interest closer to hosts of lower luminosity. The peak flux may also be more representative of hot plasma emitting UV and x-rays at much higher rates than normal. Even so, this does not fully account for the proportionally greater output at shorter wavelengths arising from a gas with a temperature greater than that of the star's surface.  The Sun's luminosity is $L_{sun}=3.8\times10^{26}$~W, and a flare peaking at 1\% of this with a  fast decay of 25 seconds would meet the  Schaefer criterion of  $10^{33}$~erg.  Slower decay rates would imply a lower peak flare emission to reach the same total energy output. 

Even without satellites constantly watching the Sun across the spectrum, we would be aware of solar flares from their nearly immediate 
effects on Earth's ionosphere, and the slower but certain arrival of ionized energetic gas and magnetic fields that alter Earth's field and create polar auroral
displays.  We know that the appearance of flares on the Sun is in part determined by two periodic factors -- the solar rotation period that returns active regions
into view on a monthly time scale, and the solar magnetic cycle that alters its field and activity on a decadal scale. For other stars, the time scales can be vastly
different,  with younger stars having faster rotation and correspondingly more powerful dynamos at work.  Much of this is understood of course, but there are several missing
pieces.   Thus stellar flares, especially those that are energetically beyond the daily modest events we see on the Sun, are an accessible laboratory for testing theories about the interaction
of magnetic fields and very energetic gases.  Detailed observations of them as well as the statistics of their occurrence for stars of different masses and evolutionary states will  add to our understanding of the extremes of flare activity on cool stars, and clarify for  the growing catalog of known planets around stars in our vicinity, which ones offer long-term stable habitability.

\section{Observations of Proxima Centauri}

Our stellar neighbor, Proxima Centauri, exhibits radial velocity variations that led to the recognition that it hosts  Proxima Centauri~b, the nearest known exoplanet~\citep{anglada2016}.  This discovery is the basis of many follow-up observations and analyses of the star and its planet, and among those  we provided several nights of monitoring with part-per-thousand precision in a collaborative photometric search for transits of other previously
undetected planets in the Proxima Centauri system~\citep{collins_false_positive_2018, blank2018, feliz2019}.  However, Proxima Centauri was long ago recognized as a flare star~\citep{shapley1951}.
It is fluctuations in flux from such stellar activity  that limit the detection of small-radius planets in habitable zones around the cool stars which are a focus of the current NASA TESS mission. 
An understanding of their variability is needed to set limits on the threshold for a signature in the signal of a true transit event.  Recently \citet{chang2018} analyzed evem more extreme hyperflares in Kepler data for M dwarf stars and noted 8 of them in the K1 lightcurve database. In the course of our study of Proxima Centauri, we found one very large flare that may qualify as a super flare and certainly sets a limit on the frequency of its more extreme flares.

\begin{figure}
  \begin{minipage}[c]{0.4\textwidth}
    \includegraphics[width=9cm]{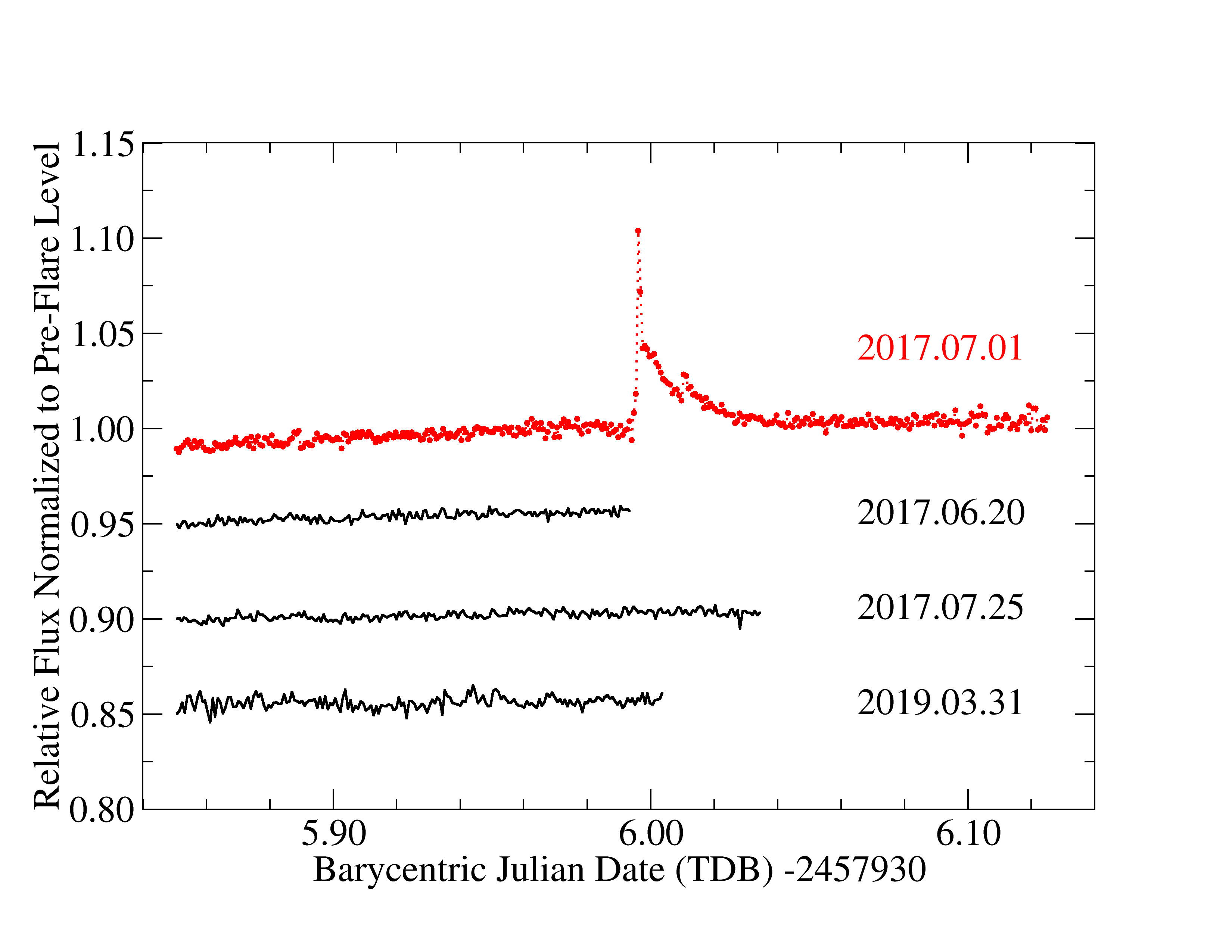}
  \end{minipage}\hfill
  \begin{minipage}[c]{0.47\textwidth} 
     \caption{The full  undetrended data set for photometry of Proxima Centauri on UTC 2017.07.01 from Mt. Kent Observatory versus barycentric Julian day.  The measurements with our CDK700 telescope were processed with AstroImageJ. Three other nights acquired with the same instrumentation are shown offset by 0.05 relative flux and by time to illustrate the normal photometry of Proxima Centauri.  \label{fig:proxima_full}}
  \end{minipage}\hfill
\end{figure}

\begin{figure}
  \begin{minipage}[c]{0.4\textwidth}
    \includegraphics[width=9cm]{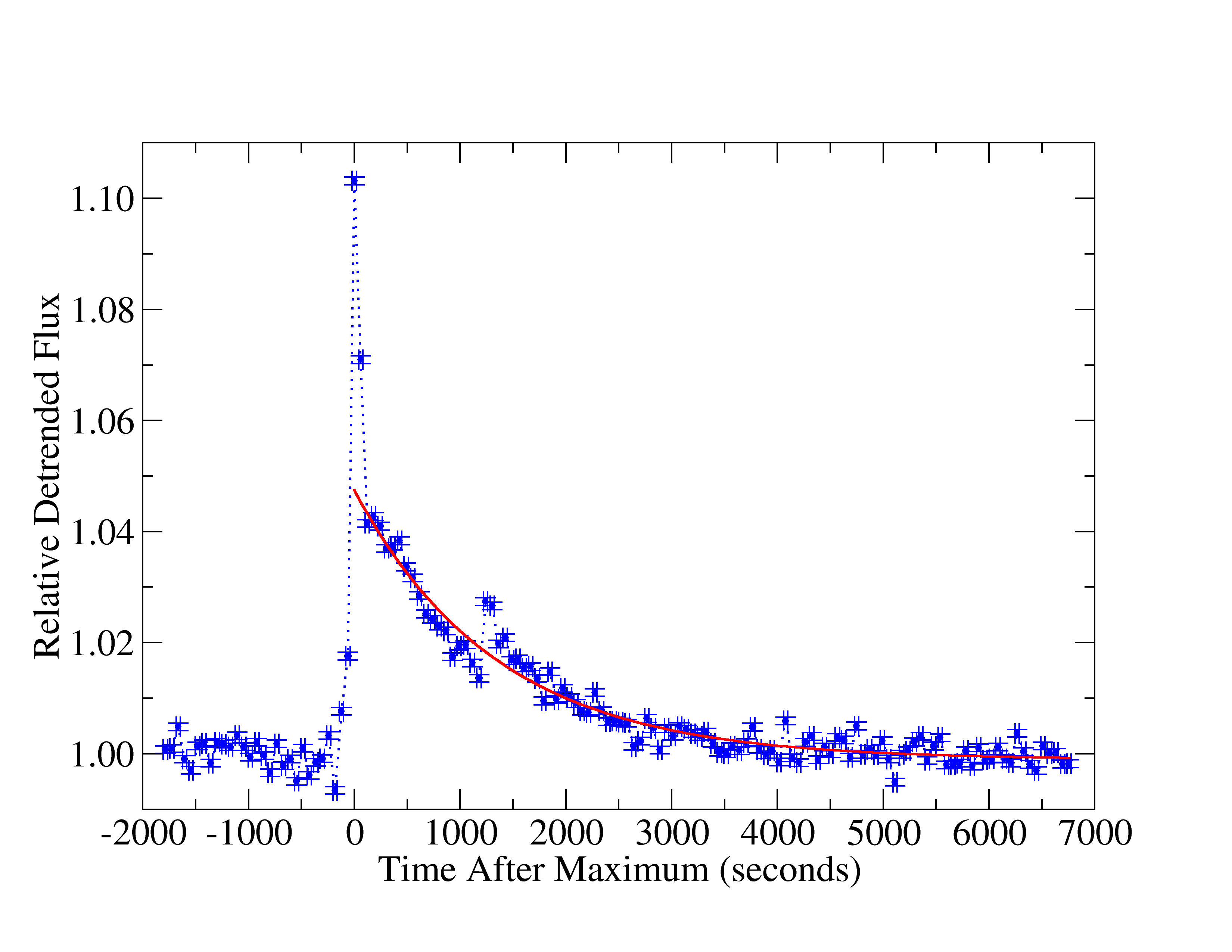}
  \end{minipage}\hfill
  \begin{minipage}[c]{0.47\textwidth} 
     \caption{Detail of the flare on  Proxima Centauri on UTC 2017.07.01. The light curve of Figure~\ref{fig:proxima_full} was detrended against airmass and is shown in seconds after the measured maximum flux. Vertical error bars are the relative flux error \citep{collins2017a}.  Horizontal bars show the exposure duration. The red line is an exponential decay model fit to the post-flare emission. \label{fig:proxima_flare} }
  \end{minipage}\hfill
\end{figure}

On  UTC 2017.07.01 at Mt. Kent Observatory in  Queensland, Australia, we acquired 385 well-guided images on a 62 second cadence with 20 second duration exposures in the Sloan $i^\prime$ band.  The instrumentation is that used for exoplanet discovery follow-up photometry. A Planewave CDK700 telescope with 0.7 meter aperture, f/6.5 corrected Dall-Kirkham optical system on a Nasmyth focus mounting combines with a Sloan filter wheel and Andor/Apogee U16 CCD (On Semi KAF-16803) to provide 0.41 arcseconds per pixel with a 27 arcminute field of view. The instrumentation is controlled by custom Linux-based software and the resulting data are processed with AstroImageJ~\citep{collins2017a} to produce light curves with noise limited primarily by photon Poisson statistics. The data shown here for Proxima Centauri were referenced to 44 other stars in the CDK700 field selected for temporal stability and freedom from blends and systematic issues.
While no transit-like events were identified that night, 
at BJD 2457935.996 there was an observed  peak  at least 10\% over pre-flare flux in that filter band.  It was followed immediately by a rapid decline trending to an exponential decay and  at least one  smaller flare event 1300 seconds later with an observed incremental peak flux increase of 1\%. Its photometry
is in Fig.~\ref{fig:proxima_full}.  Here, with a cadence limited by the transfer time
of image data from the camera, we see a partial resolution of the flare onset in the detail shown in Fig.~\ref{fig:proxima_flare}.
While the time series is not sufficiently resolved to show any fast oscillations that may occur, which would inform plasma models of details in the
physics behind the deposition of magnetic energy into hot ionized gas \citep{mclaughlin2018,jackman2019b}, it does partly resolve the onset trends and clearly shows two components to the  decay of emission.

The character of the flare event is better shown in Fig.~\ref{fig:proxima_flare} where the flux has been normalized to an interval preceding the onset of the flare and detrended for the dependence on airmass that is due to  color differences between Proxima and the other stars monitored for comparison using AstroImageJ. This figure also includes two sets of error bars.  The vertical ones are measurement-by-measurement estimates of the photon statistical error including instrumental characteristics using the AstroImageJ algorithm.~\citep{collins2017a}  The horizontal bars are not errors, but mark the 20 second exposure duration, and thus highlight the undersampling in time of the event peak. The image cadence introduces an uncertainty  in absolute timing for all measurements of $\pm0.0007$~day, while the relative timing is stable to about 1 second allowing for small variations in cadence as each image is acquired and stored.

Beginning about 1000 seconds before the peak there is a slight decline in flux exceeding the single measurement noise, accompanied by an increase in measurement-to-measurement variance. This precursor decrease is followed by an image 242 seconds prior to the flare peak that departs from this trend with a flux increase of $\approx0.5$\%. The following image shows an even greater decrease, and the image at 121 seconds prior to peak is again 0.7\% above the pre-flare baseline.  While these data suggest a pre-event pattern, they are undersampled  in time and only qualitatively  identify lightcurve features that may anticipate a large flare.  Two images later, at time zero on the plot in Fig.~\ref{fig:proxima_flare}, the observed maximum is 10.3\% above pre-flare levels.  Two images later still,  the flux is 4\% above the baseline and begins a slow decline that remains discernible for over a hour. The exposure  containing the observed peak is not time-resolved.    With an exposure duration of 20 seconds, each measurement is an  average over a rapidly changing flux when the shutter is open, and is insensitive when it is closed for the following 42 seconds of readout and processing before the next exposure.  Therefore, with an observed rapid flare onset barely resolved in two sequential 20 second exposures with a 42 second gap, it is very likely that apparent maximum is  on the decaying side of the true maximum. Furthermore, with such a rapid rate of change even if the maximum occurred within the 20 seconds the shutter was open the resulting data point is an average of the flux over that time, not a measurement of it at a precise time. 
It is also of note that since  this observation is in the Sloan $i^\prime$ band spanning 700 to 820 nm,  it does not include  spectral features such as the Ca II triplet, H$\alpha$, Na D, and Ca II H\&K that are the traditional activity indicators  enhanced in flares \citep{fuhrmeister2011}, and it may be assumed to be representative of an increase of 10\% that is broadly at least over the visible and near infrared where Proxima Centauri's spectral energy distribution is greatest. 
These data captured only  a part of the rise and structure in the fall, while not fully resolving the temporal signature of the flare's brief maximum. Therefore the use of the peak of the emission seen in the i' band as an indicator of the bolometric increase will provide a lower bound on  the actual total energy of the event.

The decay of emission seen here has two components.  The large peak falls to less than half its maximum in two image cycles, i.e. of the order of 100 seconds. It is followed by a much slower decay that is fitted well with an exponential model as shown.  That fit gives a 1/e decay time of 1350 seconds from a starting point of 4.85\% above the baseline and can be traced to   5000 seconds after the event where it disappears into the noise.  On this, there is a second flare occurring at 1302 seconds which is 2.7\% above the baseline and approximately 1\% above the exponential model.  The long duration of the decay contributes significantly to the total energy in the event, and integrates to $63 L_\star$, where   Proxima's bolometric luminosity is  $L_\star = 0.00151 \pm 0.00008 L_\odot$~\citep{ribas2017}. The contribution from the fast peak component can only be approximated since it changes so rapidly it is undersampled.  A minimal estimate would be  $10 L_\star$, significantly less than the energy in the long tail.  However, if the event peak has a blackbody temperature well above that of the cool star as is likely, this will be orders of magnitude too small. For example, a model of the flare spectral energy distribution developed by \citet{davenport2012} shows that most of the energy is emitted in the ultraviolet, and that the spectrum decreases in flux through the visible into the near-infrared.   Certainly the UV and X-ray emission that makes the flare a hazard for habitability occurs in this brief time. At a minimum, not allowing for the spectral differences in the two temporal components, the integrated flare energy is greater than  $10^2 L_\star$  or $6\times10^{32}$ ergs.
Using a criterion based on fractional peak emission relative to the ambient star luminosity of more than 1\%, this event  is easily within the superflare class. It also easily exceeds  the $10^{33}$~erg energy threshold as a superflare with allowance for a higher temperature at the peak of the emission. 
An exact calibration of a Sloan i'-band observation such as ours to the bolometric flux of a rare superflare event is not possible without a comprehensive model of the flare physics, a validated spectral energy distribution for flares of this type, and at least a fitting of the observed passband to the predicted emission with that model. Some work in that direction has been done, and  \citet{cram1982} have investigated models for stellar flares  to test predictions of flare spectra.  From that work it appears that the hydrogen Balmer emission that is often associated with flares (see the observation of  \citet{carter1988}, for example) arises in a separate region from the white light continuum emission.  Thus observations in a band that include the H$\alpha$ in the red may not be representative the the broad spectrum or bolometric flux without allowance for the Balmer component. Similarly, it is not appropriate to assume that the flare spectrum is like that of the quiescent star, and, as the observers of the Carrington event noted,  what we regard as a superflare today has the appearance of a hot star, suggesting an effective temperature of greater than $10^4$~K. \citet{davenport2012} modeled the spectrum with a UV component to represent this effect.  We would therefore interpret the data in the Sloan-i' band as an observation of a quasi-blackbody in the Rayleigh-Jeans regime where there is a $\lambda^{-4}$ dependence of the spectral flux. If that is the case, then the longer wavelength 800~nm  measurements will see a flux increase that is of the order of 15\% of the increase seen in a bandpass centered on 500~nm.  

\section{Other Observations and Habitability Implications}

The spectral energy distribution the flare compared the average emission of the star determines the relative flux increase, affecting both the detectability of the flare by distant observers, and  habitability of a planet at equilibrium with the quiescent stellar state.  An exceptional example is the recently detected white light flare on an L2.5 cool dwarf star, where a flux increase of 10 magnitudes in the V-band was seen.~\citep{jackman2019a}. The calculated flare energy, based on an assumed flare temperature of 9000~K,  for that case was $3.4\times10^{33}$~erg. The flare's time dependence was similar to what we observe for the event on Proxima Centauri with an unresolved onset and peak, followed by an exponential-like decay back to the quiescent background with a time constant of less than 5 minutes.  

The occurrence rate of flares on Proxima Centauri based on a census of space-based data from the Microvariability and Oscillations of STars (MOST) satellite has been estimated by \citet{davenport2016}. The MOST filter curve includes the visible and near-infrared spectrum with spectral features that are sensitive to flare events.  Within that passband, the star was found to have flares producing flux increases of more than 0.5\% at a mean rate of 63 per day, and superflares at a rate of 8 per year.  Several were noted in the satellite database, with 63 second cadence data and flux increases comparable to our ground-based observation.  The largest flare highlighted in the MOST data occurred at HJD 2451545 + 5611.246 (see their Fig.~1) with a peak amplitude increase of 26\% above the baseline and a smooth decay back to the baseline within 0.02 day. As in other detections of rare strong flares, the MOST data do not resolve the onset.  The flare energy measurement was made by comparing to the star's quiescent luminosity within the MOST filter band. \citet{howard2018} reported a singularly strong flare occurred in UT 2016.03.18, and many other weaker ones,  detected by the Evryscope survey.     Another  massive flare on Proxima was observed with the Atacama Large Millimeter/submillimeter Array (ALMA) on UT 2017.03.24. The observed event lasted for  about 1 minute during which the peak flux spectral density was 1000 times that of the Sun's comparable quiescent emission in the ALMA band at 233 GHz~\citep{macgregor2018}. 

\citet{shibayama2013} have examined the statistical properties of superflares on solar-type stars observed with Kepler, searching for flares on G-type stars in the database.  They find an occurrence rate of $\sim10^{-3}$ yr$^{-1}$, while some outliers have occurrence rates of greater than 1 yr$^{-1}$ that could be attributed to extremely large starspots.  The time dependence of the flares in this study with peak energies of $\sim10^{36}$~erg was also generally not resolved at the flare onset, and had an exponential-like decay with a time constant of less than 0.2 day and with multiple post-primary-flare peaks. Also for comparison with well-observed solar flares, they use a total energy of $10^{32}$ erg for an X10 class flare that sets the bar for the Carrington event and for superflares seen in other stars to be above the level of exceptional solar events 

\citet{segura2010} have analyzed the possible effects of the UV emission from a flare of this magnitude on the atmosphere of an Earth-like planet in the habitable zone of an M-dwarf star AD Leo as a proxy for other stars such as Proxima Centauri. They conclude that such flares may not be a hazard for life on the surface of the planet because the atmosphere recovers on time scales of years.  Of course, repeated flaring events may affect that recovery, and the one we observe fits in the class of superflares which \citet{lingam2017a} say would jeopardize life in the habitable zone of a planet orbiting a star such as this.  

\citet{ohare2019} measured radionuclides in Greenland ice cores and found evidence of a previously unknown solar proton event 2610 years before the present  that was of the same order of strength  as one identified in the AD 774/775 time frame, while the Carrington event is not found in most ice core nitrate records~\citep{wolff2012}. The possibility of solar superflares occurring on time scales of $>10^3$ years remains open.


\section{Conclusions}

We have observed a flare on Proxima Centauri resulting in  a peak increase in  near-infrared continuum luminosity of more than 10\%  and a decay with two components.  The first is rapid and within 100 seconds the emitted flux is down by half.  The second is much slower, with an exponential decay constant of 1350 seconds.  The event fits in the class of rare superflares  which some analyses say would jeopardize life on a planet  in the star's habitable zone, especially if the interval between comparable flares is less than the atmospheric recovery time. The statistics of such events on Proxima Centauri and other M-dwarf stars is currently under study.  Nevertheless, available data suggest rates of several per year for events with sharply peaked emission and exponential decays lasting several minutes.   The initiation of these flares that occurs within a few seconds is quite brief.  Within the duration of a typical astronomical photometric observation, the peak of the flare flash is lessened,  and similar events may  have gone largely undetected on this and other cool stars until recently. Nevertheless,  a rate of a few events per year is high enough that occasional time series ground-based observations will see one such as we recorded.  Much more frequent lower energy flares  appear as an increase in noise versus the desired signal of transit-like events in photometric data from NASA TESS where the fastest cadence snapshots are on 2 minute intervals and full frame images are integrations over 30 minutes~\citep{ricker2015,oelkers2018}. Thus the high precision and low intrinsic noise in the TESS data for M-dwarf stars may define flare rates and the probability of superflare events on the cool stars that are the TESS primary targets for detecting habitable planets similar to Earth.  Higher cadence measurements of flares both from the ground in the visible and near-infrared, and from space to achieve full spectral coverage, is essential to understand the physical processes that are at work in those stars that may host habitable planets. Because the flux increases seen here and elsewhere are a large fraction of the stellar flux, observations optimized for stellar flare measurements may be different from those optimized for the detection of shallow transits, and should include several passbands to establish an effective flare temperature as well as bolometric luminosity, precursor behavior, and post-flare decay.

\section*{Acknowledgements}
We are grateful to the KELT followup team organization, and to Karen Collins for coordinating the observations of Proxima Centauri that led to the discovery reported here.


\bibliographystyle{mnras}
\bibliography{proxima_flare}

\bsp	
\label{lastpage}
\end{document}